\documentclass{emulateapj}

\usepackage{graphicx}
\usepackage{amsmath}
\newcommand{\DeltaQ}{\Delta Q}
\begin{document}
%


\title{Constraining neutron star matter with QCD}

\hfill CERN-PH-TH/2014-032, HIP-2014-02/TH

\author{Aleksi Kurkela$^1$, Eduardo S. Fraga$^{2,3,4}$, 
J\"urgen Schaffner-Bielich$^{2}$, and Aleksi Vuorinen$^5$}

\affil{$^1$Physics Department, Theory Unit, CERN, CH-1211 Gen\`eve 23, Switzerland}
\affil{$^2$Institute for Theoretical Physics, Goethe University, D-60438 Frankfurt am Main, Germany}
\affil{$^3$Frankfurt Institute for Advanced Studies, Goethe University,
  D-60438 Frankfurt am Main, Germany} 
\affil{$^4$Instituto de F\'\i sica, Universidade Federal do Rio de Janeiro,
Caixa Postal 68528, 21941-972, Rio de Janeiro, RJ, Brazil}
\affil{$^5$Department of Physics and Helsinki Institute of Physics, P.~O.~Box
  64, FI-00014 University of Helsinki, Finland} 


\begin{abstract}  
	In recent years, there have been several successful attempts to constrain the equation of state of neutron star matter using input from low-energy nuclear physics and observational data. We demonstrate that significant further restrictions can be placed by additionally requiring the pressure to approach that of deconfined quark matter at high densities. Remarkably, the new constraints turn out to be highly insensitive to the amount --- or even presence --- of quark matter	inside the stars.

\end{abstract}




\section{Introduction}
\label{sec:intro}

The equation of state (EoS) of cold and dense strongly interacting matter, which determines the inner structure of compact stars
\citep{Glendenning:1997wn}, is encoded in its fundamental theory, Quantum Chromodynamics (QCD). A full nonperturbative determination of the pressure of the theory is still out of reach due to the so-called Sign Problem of lattice QCD \citep{deForcrand:2010ys}. Nevertheless, the methods of chiral effective field theory (EFT) of nuclear forces \citep{Epelbaum:2008ga} and high-density perturbative QCD (pQCD) \citep{Kraemmer:2003gd} have matured enough to provide reliable predictions for the EoS in the limits of low density nuclear matter and dense quark matter, respectively. In particular, by now both approaches produce results with reliable error estimates, implying that it is finally possible to quantitatively estimate our understanding of the neutron star matter EoS. 

During the past couple of years, several articles have addressed the
determination of the neutron star EoS by combining insights from
low-energy chiral EFT with the requirement that the resulting EoSs support the most massive stars observed (see e.g.~\cite{Hebeler:2013nza}). In particular, the discovery of neutron stars with masses around two solar masses \citep{Demorest:2010bx,Antoniadis:2013pzd} has recently been seen to lead to strong constraints on the properties of stellar matter
\citep{Lattimer:2012nd}. While otherwise impressive, these analyses have solely concentrated on the low density regime, and have typically applied no microphysical constraints beyond the nuclear saturation density $n_0$. This has resulted in EoSs that behave very differently from that of deconfined quark matter even at rather high energy densities. 

In the present work, our aim is to demonstrate that the EoS of neutron star matter can be significantly further constrained by requiring it to approach the quark matter one at high density. To do this, we use the state-of-the-art result of \cite{Fraga:2013qra}, where a compact expression for the three-loop pressure of unpaired quark matter, taking into account the nonzero value of the strange quark mass, was derived (see also \cite{Kurkela:2009gj} and \cite{Kurkela:2010yk} for details of the original pQCD calculation). A particularly powerful outcome of the analysis is that the high density constraint significantly reduces the uncertainty band of the stellar matter EoS even at low densities, well below a possible phase transition to deconfined quark matter. This implies that the $M$-$R$ relations we obtain are more restrictive than previous ones even for pure neutron stars.

\begin{figure}
\includegraphics[width=0.45\textwidth]{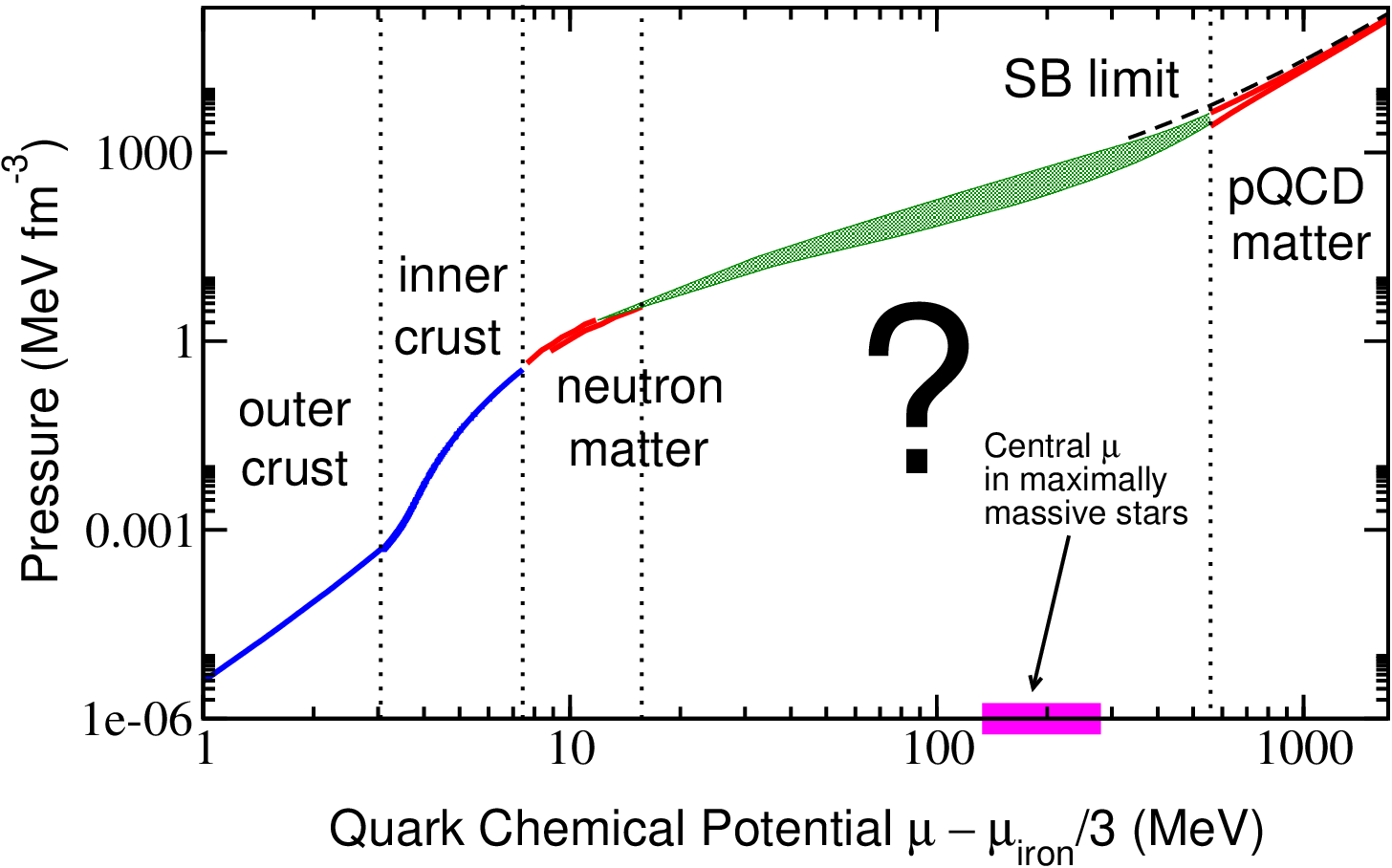}
\caption{Known limits of the stellar EoS on a logarithmic scale. On the horizontal axis we have the quark chemical potential (with an offset so that the variable acquires the value $0$ for pressureless nuclear matter), and on the vertical axis the pressure. The band in the region around the question mark corresponds to the interpolating polytropic EoS that will be introduced in this work.} 
\label{fig1}
\end{figure}

In practice, our calculation proceeds as follows (see also fig.~\ref{fig1}): at densities below $1.1n_0$, we employ the chiral EFT EoS of \cite{Tews:2012fj}, assuming the true result to lie within the error band given in this reference. At baryon chemical potentials above 2.6 GeV, where the relative uncertainty of the quark matter EoS is as large as the nuclear matter one at $n=1.1 n_0$, we on the other hand use the result of \cite{Fraga:2013qra} and its respective error estimate. Between these two regions, we assume that the EoS is well approximated by an interpolating polytrope built from two ``monotropes'' of the form $P(n)$=$ \kappa n^\Gamma$. These functions are first matched together in a smooth way, but later we also consider the scenario of a first-order phase transition, allowing the density to jump at the matching point of the two monotropes.

Varying the polytropic parameters and the transition density over ranges limited only by causality, we obtain a band of EoSs that can be further constrained by the requirement that the EoS support a two solar mass star. This results in EoS and $M$-$R$ bands that are significantly narrower than ones obtained without the high density constraint (see Fig.~\ref{fig10}). An important check of the robustness of our construction is that the obtained band is largely unaffected by the nature of the assumed phase transition or by the introduction of a third interpolating monotrope.

\begin{figure}
\includegraphics[width=0.45\textwidth]{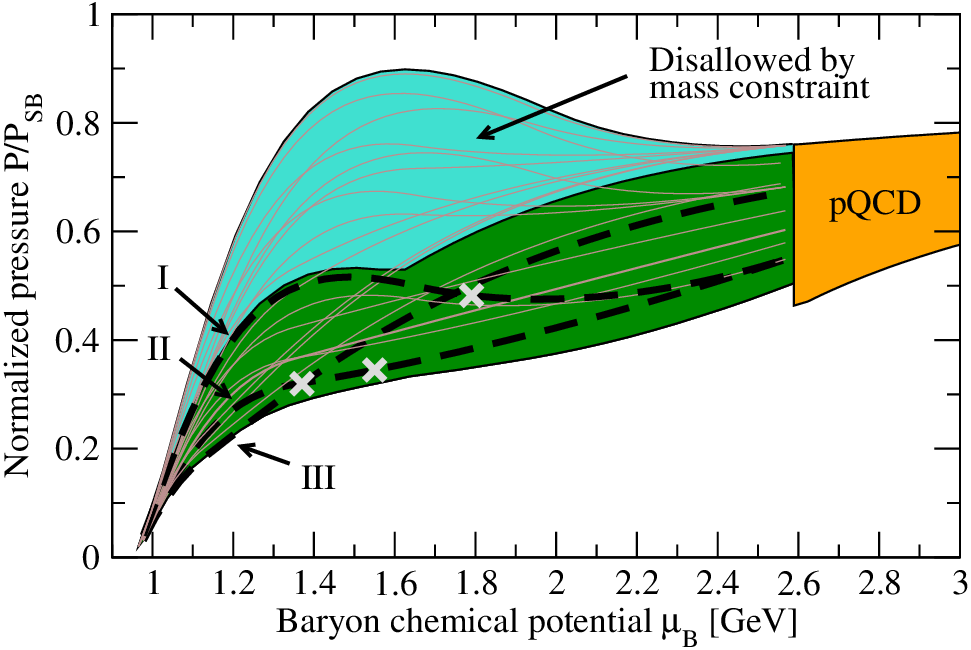}
\caption{The interpolated pressure of nuclear and quark matter, normalized by the pressure of a gas of free quarks and shown together with the pQCD result at high densities. All generated EoSs lie within the shaded green and turquoise areas, of which only the green ones support a star of $M=2M_{\odot}$. Three representative EoSs marked with I-III have crosses denoting the maximal chemical potential reached at the center of the star.}
\label{fig2}
\end{figure}

Clearly, both our setup and results bear some resemblance to those of \cite{Hebeler:2013nza}. An important difference between these two calculations is, however, that while the authors of the latter paper imposed a set of somewhat \textit{ad hoc} constraints on their polytropic parameters, for us this is not necessary, as the high density constraint automatically restricts these numbers. In this vein, one can in fact argue that our calculation gives an \textit{a posteriori} justification for many of the choices made in \cite{Hebeler:2013nza}. At the same time, it is also important to note that the ease with which we are able to perform the matching between the low and high density EoSs --- and the high maximum masses we obtain for the stars --- is in stark contrast with many earlier attempts to directly match nuclear and quark matter EoSs onto each other (cf.~e.g.~\cite{Fraga:2001xc} and \cite{Alford:2004pf}).

Our paper is organized as follows. In section 2, we first explain the details of our calculation, i.e.~introduce the low and high density EoSs used as well as the specific parametrization of our interpolating polytropes. After this, we proceed to display and analyze our results in section 3, covering both the EoSs, $M$-$R$ relations, and various correlations between the parameters appearing in our calculation. In section 4, we finally draw our conclusions.

\section{Methodology}
\label{sec:method}

\subsection{Low-density EoS from chiral EFT}

Outside the dense inner core of a compact star, one expects to find somewhat more dilute nuclear matter. The composition of this medium ranges from a gas of nuclei (inside an electron sea) in the outer stellar crust to increasingly neutron-rich matter in the inner crust and outer core of the star. The EoS of the latter type of matter has traditionally been estimated through many-body calculations employing phenomenological potentials, typically accounting for two- and three-nucleon interactions (see e.g.~\cite{Akmal:1998cf}). More recently, developments in chiral effective theory (EFT) have, however, significantly systematized this procedure and in particular provided a formal basis for the hierarchy amongst contributions coming from different types of interactions
\citep{Coraggio:2012,Gandolfi:2012,Hebeler:2009iv,Holt:2012,Sammarruca:2012,Tews:2012fj}.

\begin{figure}
\includegraphics[width=0.45\textwidth]{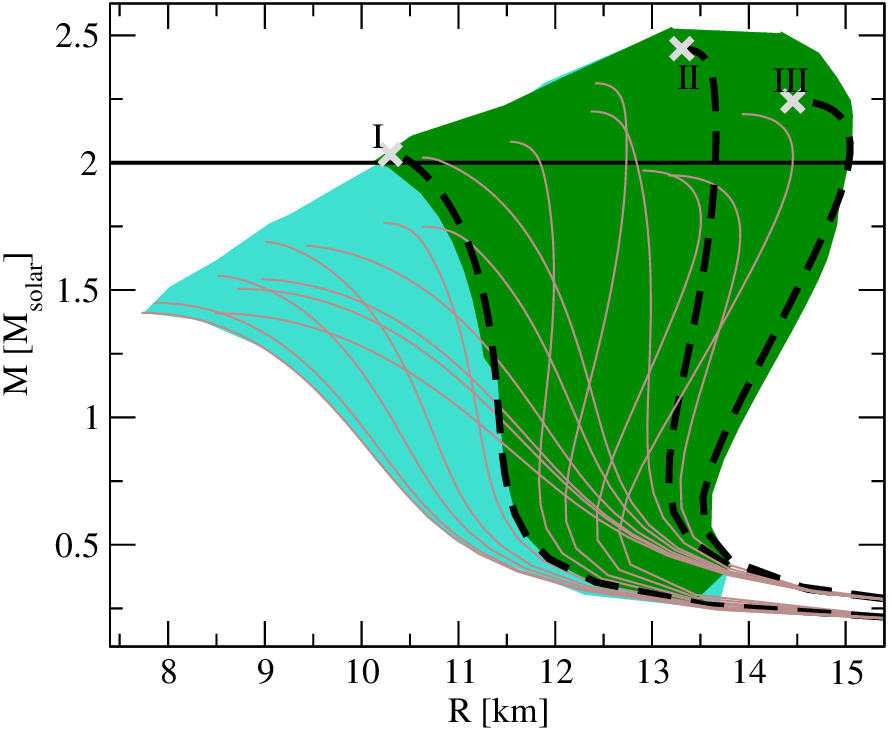}
\caption{Two $M-R$ clouds composed of the EoSs displayed in
  Fig.~\ref{fig2}. The color coding is the same as there, as is our notation for the three representative EoSs I-III.} 
\label{fig3}
\end{figure}

At the moment, calculations within chiral EFT \citep{Epelbaum:2008ga} have reached a state, where uncertainties related to the details of many-body simulations are negligible, and nearly the entire remaining error in the EoS originates from the determination of various coupling constants of the EFT itself. The magnitude of these uncertainties grows rapidly with density, such that at nuclear saturation density the pressure of neutron star matter is currently known to roughly $\pm 20\%$ accuracy \citep{Tews:2012fj}. In the near future,
it is expected that these uncertainties will further decrease through the emergence of more precise constraints for the low-energy couplings of the chiral EFT and the incorporation of higher-order interactions in the theory. See e.g.~\cite{Hebeler:2013nza} for a more systematic analysis of these issues.

The current state of the art in the determination of the nuclear matter EoS can be found in \cite{Tews:2012fj}, the results of which are heavily based on the earlier work of \cite{Hebeler:2009iv}. In \cite{Hebeler:2013nza}, these EoSs are given in a tabulated form for densities $n\in[0.6,1.1]\,n_0$, with $n_0$ the nuclear saturation density. In our calculations, we use both the maximally soft and stiff variations of this nuclear EoS, corresponding to the lower and upper limit of the pressure at a given number density. For these two results, also the dependence of the baryon chemical potential on the density is somewhat different. For densities below 0.6$n_0$, we in addition need the crust EoS, which can be found e.g.~from \cite{Baym:1971,Negele:1971vb,Ruster:2006}.

\begin{figure}
\includegraphics[width=0.45\textwidth]{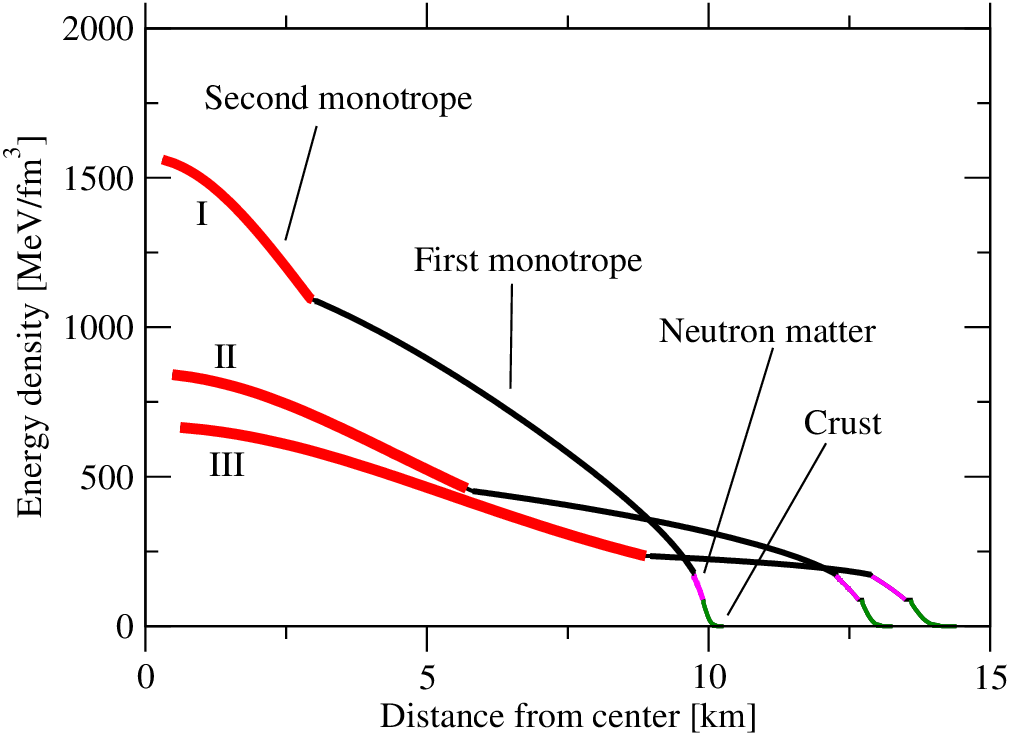}
\caption{The internal structure of the maximally massive stars corresponding to the three EoSs I-III of Table \ref{table1}. The energy densities are all continuous due to the smoothness of the matching procedure (no first order phase transition) in these cases.}
\label{fig4}
\end{figure}

\subsection{High-density EoS from pQCD}

The EoS of cold quark matter is accessible through perturbative QCD at high densities, and has indeed been determined to order $\alpha_s^2$ in the strong coupling constant. This calculation was first carried out at vanishing quark masses in \cite{Freedman:1976ub,Baluni:1977ms}
(cf.~also~\cite{Blaizot:2000fc,Fraga:2001id,Andersen:2002jz,Vuorinen:2003fs}), and later generalized to systematically include the effects of a nonzero strange quark mass at two \citep{Fraga:2004gz} and three loops \citep{Kurkela:2009gj}.

Like all perturbative results evaluated to a finite order in the
coupling, also the quark matter EoS is a function of an unphysical parameter, the scale of the chosen renormalization scheme (here modified minimal subtraction) $\bar{\Lambda}$. This dependence, which diminishes order by order in perturbation theory, offers a convenient way to estimate the contribution of the remaining, undetermined orders, and thus serves as quantitative measure of the inherent uncertainty in the result.

In a recent study of \cite{Fraga:2013qra}, it was demonstrated that the complicated numerical EoS derived in \cite{Kurkela:2009gj} can be cast in the form of a simple fitting function for the pressure in terms of the baryon chemical potential $\mu_B$. Fixing the strong coupling constant and the strange quark mass at arbitrary reference scales (using lattice and experimental data), the EoS of quark matter in $\beta$-equilibrium assumes the form
\begin{align}
P_{\rm{QCD}}(\mu_B) &= P_{\rm{SB}}(\mu_B) \left( c_1 - \frac{a(X)}{(\mu_B/{\rm GeV}) - b(X)} \right), \label{eq:pressure}\\
a(X) &= d_1 X^{-\nu_1},\quad
b(X) = d_2 X^{-\nu_2},
\end{align}
where we have denoted the pressure of three massless noninteracting quark flavors by
\begin{align}
P_{\rm SB} = \frac{3}{4\pi^2}(\mu_B/3)^4.
\end{align}

The parameters of the above EoS can be shown to acquire the optimal values (see \cite{Fraga:2013qra} for details) 
\begin{eqnarray}
c_1=0.9008 \quad & d_1= 0.5034 &\quad d_2 = 1.452 \\
\nu_1 &= 0.3553 \quad \nu_2&= 0.9101,
\end{eqnarray}
while the dependence of the result on the renormalization scale is encoded in the functions $a(X)$ and $b(X)$. They in turn depend on a dimensionless parameter proportional to $\bar{\Lambda}$, $X\equiv 3\bar{\Lambda}/\mu_B$, the value of which is let to vary from $1$ to $4$. The resulting expression can be seen to correctly reproduce the full three-loop pressure, quark number density and speed of sound to per cent accuracy for baryon chemical potentials smaller than $6$ GeV.

\subsection{Polytropes and their matching}

\begin{figure}
\includegraphics[width=0.45\textwidth]{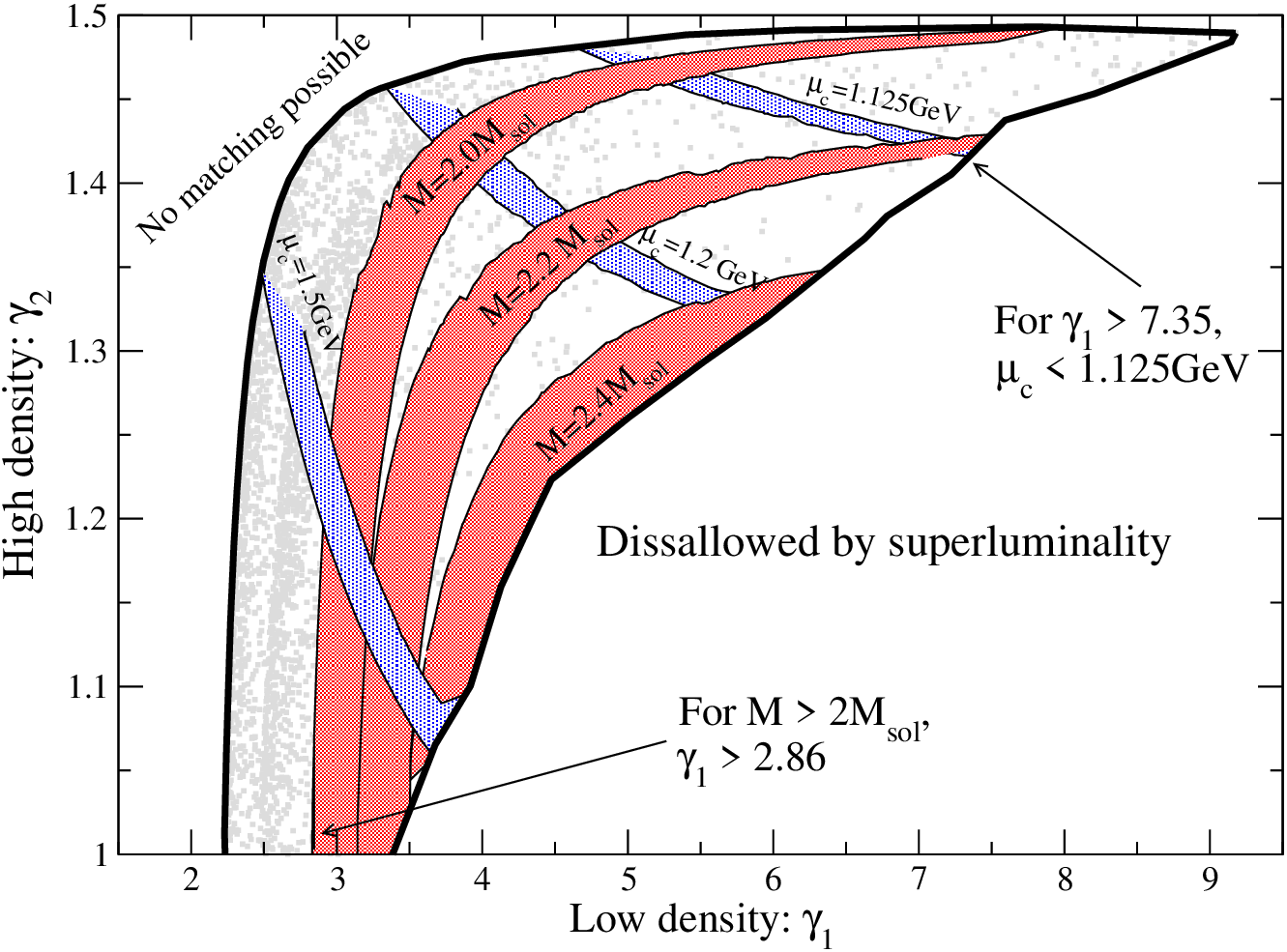}
\caption{The maximal star mass and $\mu_c$ as functions of the polytropic indices $\gamma_1$ and $\gamma_2$. The red bands corresponding to $M=2$, $2.2$ and $2.4M_{\odot}$ denote the ranges of $\gamma$s, for which stars of these masses can be obtained, while the widths of the bands reflect the uncertainties in the low and high density EoSs. The white region in the lower right hand corner is excluded due to superluminality (which can occur at a density higher than reached at the center of the star). The upper left hand corner would on the other hand require $X$ to be larger than 4, and is thus disallowed by our pQCD EoS. The gray dots finally represent our 3500 randomly generated EoSs.}
\label{fig5}
\end{figure}

At a baryon density $n^\text{begin}\equiv 1.1 n_0$, the chiral EFT prediction for the EoS of neutron-rich nuclear matter has an uncertainty of $\pm 24\%$ --- an accuracy matched by the perturbative quark matter pressure at $\mu_B^\text{end}\equiv 2.6$ GeV. To parameterize the (unknown) behavior of the EoS between these two limits, a natural choice is to employ one or more monotropes of the form $P_i(n)= \kappa_i n^{\gamma_i}$, matched together at a set of intermediate chemical potentials. As using a single monotrope is seen to lead to an overconstrained system, and the use of more than two monotropes has only a minor effect on the results, in most of our forthcoming analysis our interpolating EoS is composed of exactly two monotropes. In the beginning we assume that the matching of the two monotropes is smooth, and that there is no jump in the number density at the matching point. This assumption is, however, relaxed later, when we study the scenario of a first-order phase transition between the nuclear and 
quark matter phases.

\begin{figure}
\includegraphics[width=0.45\textwidth]{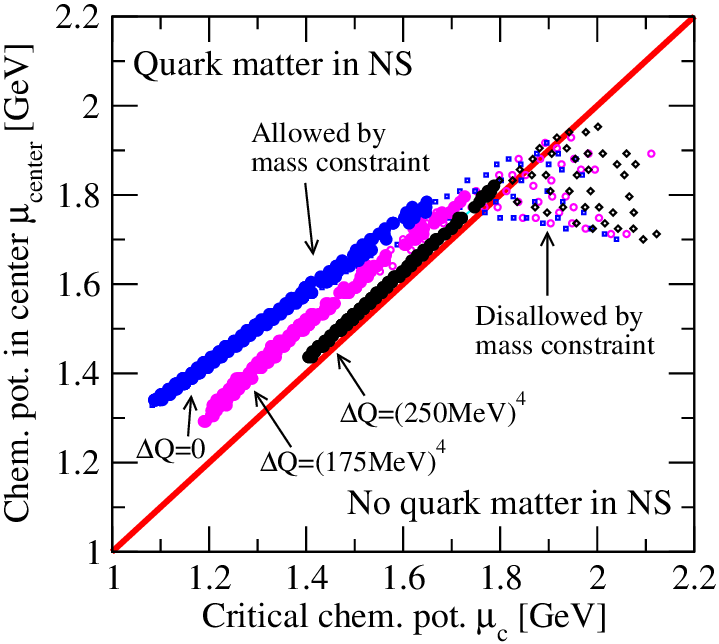}
\caption{The maximal chemical potential reached at the center of a maximum mass star as a function of the critical chemical potential. We display here results corresponding to several values of the parameter $\Delta Q$, standing for the strength of the phase transition: $\DeltaQ=0$ (blue), $\DeltaQ=(175\rm{MeV})^4$ (magenta), and $\DeltaQ=(250\rm{MeV})^4$ (black). The open points correspond to EoSs that cannot support a $M=2M_{\odot}$ star, while the solid
points are allowed by the mass constraint.}  
\label{fig6}
\end{figure}

Concretely, our calculation proceeds as follows. In both intermediate
intervals (to be specified later), we define 
\begin{eqnarray}
P_i(n)&=& \kappa_i n^{\gamma_i}, \;\;\; i=1,2,
\end{eqnarray}
or equivalently
\begin{eqnarray}
P_i(\mu_B)&=&
\kappa_i \left(n_i^{\gamma_i-1} + \frac{\gamma_i-1}{\kappa_i \gamma_i}(\mu_B-\mu_{B,i}) \right)^\frac{\gamma_i}{\gamma_i-1},
\end{eqnarray}
where $\mu_{B,i}$ and $n_i$ stand for the baryon chemical potential and baryon
density at the lower edge of the interval. The matching procedure then
consists of the following sequential steps:
\begin{enumerate}
\item Below $n^\text{begin}$, we employ either the soft or stiff nuclear matter EoS of \cite{Tews:2012fj}. This provides us with the corresponding values for $\mu_B$ and $P$ at this point (for the soft and stiff EoS, respectively),
\begin{eqnarray}
\mu_{B}= 0.9775\textrm{GeV}  & \quad P = 3.542 \textrm{MeV/fm}^3,  \label{Nucl:s}\\
\mu_{B}= 0.9657 \textrm{GeV}& \quad  P  = 2.163 \textrm{MeV/fm}^3,\label{Nucl:h}
\end{eqnarray}
as well as the parameters $\mu_{B,1}$, $n_1$ and $\kappa_1$ for the first monotrope.
\item Choosing a (positive) value for $\gamma_1$, we evolve the first  monotrope until the matching point $\mu_c$. Here, we then use this function to obtain the initial data $n_2$ and $\kappa_2$ for the second one (recalling also that $\mu_{B,2}=\mu_c$). If we wish to have a first-order phase transition at this point, we furthermore add an order ${\mathcal O} (\Lambda_\text{QCD}^4/\mu_B)$ contribution to $n_2$.
\item We evolve the second monotrope until $\mu_B=\mu_B^\text{end}$, evaluate the values of $p$ and $n$ there, and try to find an $X\in [1,4]$, for which smooth matching to the quark matter EoS is possible. If no such value is found, the EoS is discarded; otherwise the corresponding quark matter EoS takes over beyond this point.
\item Finally, we evaluate the speed of sound $c_s$ of our EoS over the entire interval from $n=n^\text{begin}$ to $\mu_B=\mu_B^\text{end}$, and locate its maximum there. Should this value exceed 1, the EoS is again discarded as superluminal.
\end{enumerate}

\section{Results}
\label{sec:beyond}
\subsection{Bitropic interpolation}

The outcome of the matching and interpolation procedure explained in the previous section is displayed in Fig.~\ref{fig2} for the case of two smoothly matched monotropes. Here, the two bands (together) correspond to a set of 3500 physical EoSs that were constructed from flat probability distributions in both $\mu_c\in[1.05, 2.4]$ GeV and
$X\in[1,4]$. The corresponding polytropic indices are seen to vary over the intervals $\gamma_1\in [2.23, 9.2]$ and $\gamma_2\in [1.0, 1.5]$, while the matching point $\mu_c$ resides between $1.08$ and $2.05$ GeV; the tight constraint for $\gamma_2$ clearly originates from the matching to the pQCD pressure. Alongside with the bands, we also show a selected set of representative EoSs, listed in Table~\ref{table1}, of which three are marked in bold and tabulated in Tables~\ref{table2}--\ref{table4}. The typical structure of the EoSs is such that the maximal stiffness (or $c_s^2$) is reached just below $\mu_c$.

\begin{figure}
\includegraphics[width=0.45\textwidth]{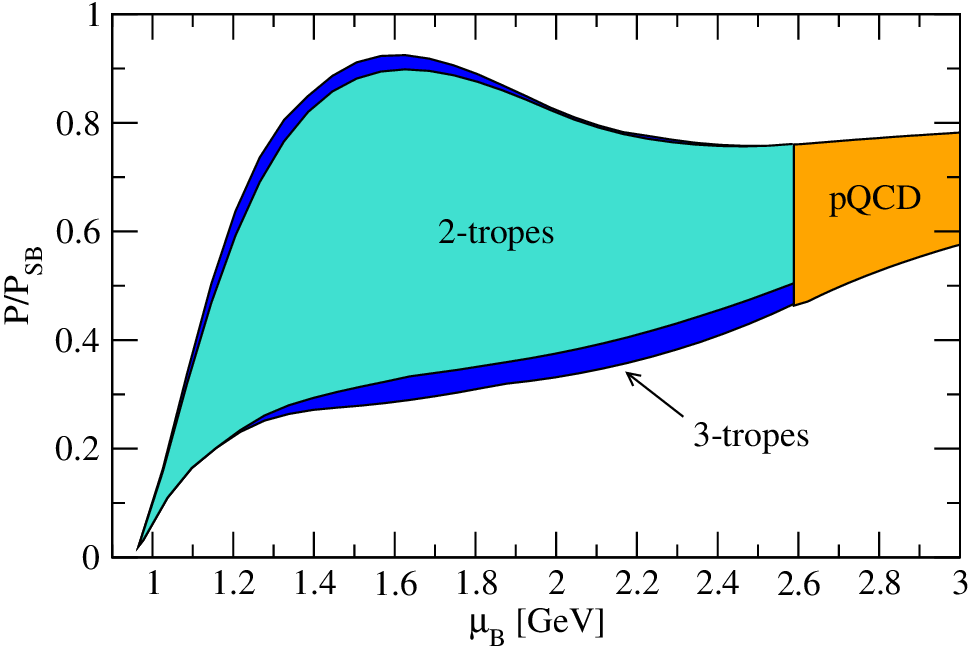}
\caption{The change in the EoS cloud (without a mass constraint) when allowing for a third interpolating monotrope between the low and high density regimes.} 
\label{fig7}
\end{figure}

In Fig.~\ref{fig3}, we next display two clouds of $M-R$ curves corresponding to all of our generated EoSs. The maximal masses of the stars fall inside the interval $M_{\rm max}\in [1.4, 2.5]\,M_{\odot}$, while their radii lie in the range $R\in [8,14]$ km. At the same time, the maximal chemical potentials encountered at the center of the star satisfy $\mu_{\rm center} \in [1.33,1.84]$ GeV, corresponding to maximal central densities of $n\in[3.7,14.3]\, n_0$. This falls right in the middle of the interval between the nucleonic and pQCD regions, where the EoS is equally constrained by its low and high density limits. In addition, we show here a number of individual $M$-$R$ curves, corresponding to the 20 EoSs listed in Table~\ref{table1}, and mark the maximal chemical potentials of the three special EoSs of Fig.~\ref{fig2} (I, II and III) with crosses. For these three cases, Fig.~\ref{fig4} additionally displays the internal structure of the maximally massive stars; here, the softening of the EoS when 
approaching the perturbative densities is seen as a faster growth of the energy density near the center of the star.

The stellar matter EoS can of course be further constrained by demanding that it is able to support the observed two solar mass star. The effects of this constraint on the EoS and $M-R$ clouds of Figs.~\ref{fig2} and \ref{fig3} are visible as the dark green areas. In particular, we find that remarkably {\it with the additional mass constraint the relative uncertainty in the EoS is reduced to less than $\pm 30\%$ at all densities}. For these EoSs, the maximal chemical potentials are bound from above by $\mu_{\rm center}<1.77$ GeV, and the central densities by $n<8.0\, n_0$.

From Fig.~\ref{fig3}, one can in addition read that for $1.4M_\odot$ neutron stars, our allowed radii range between 11 and 14.5 km, while for $2M_\odot$ pulsars, $R\in [10,15]\,$km. It is also worth noting that within the bitrope approach, we find no configurations with masses above $2.5M_\odot$. In comparison with the findings of \cite{Hebeler:2013nza}, our upper and lower limits for the radii are consistently larger by about $1$ km, while our most massive configurations are lighter by about $0.5M_{\odot}$.

Moving on to an analysis of the polytropic indices $\gamma_1$ and $\gamma_2$, we display in Fig.~\ref{fig5} a contour plot of the maximal star mass as a function of these parameters. As larger $\gamma_1$ translates into a stiffer equation of state below $\mu_c$, it is natural that requiring the reaching of a given mass sets a lower bound for $\gamma_1$. In particular, reaching a two solar mass star translates to the condition $\gamma_1 > 2.86$, while
$M>2.4M_{\odot}$ translates to $\gamma_1> 3.5$.

Polytropic EoSs with the index $\gamma$ larger than 2 become eventually superluminal, implying that the larger $\gamma_1$ is, the smaller the value of $\mu_c$ has to be in order for the EoS to stay subluminal. The blue bands in Fig.~\ref{fig5}, which stand for constant values of $\mu_c$, demonstrate this fact: for $\gamma_1= 7.35$, we must require $\mu_c < 1.125$, while for $\gamma_1= 5.9$ we get $\mu_c < 1.2$ GeV, and for $\gamma_1=3.65$ simply $\mu_c<1.5$ GeV. As the second segment of the polytrope is typically significantly softer than the first, the stars quickly become unstable once their center reaches $\mu_c$. The maximal $\mu_B$'s are thus typically only slightly larger than the corresponding $\mu_c$'s, as one can see from Fig.~\ref{fig6}.

\begin{figure}
\includegraphics[width=0.45\textwidth]{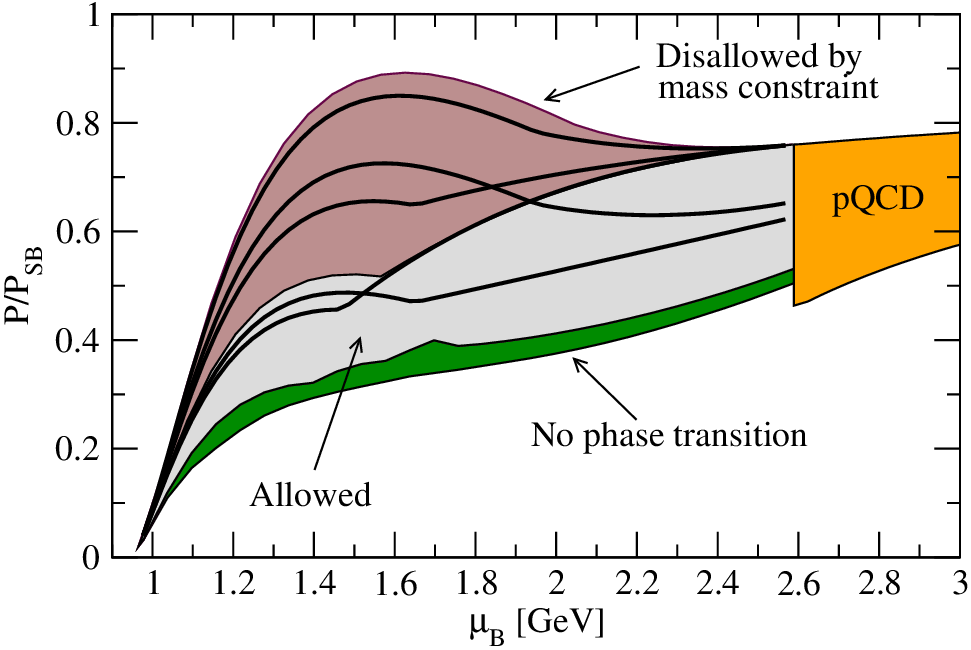}
\caption{The EoS cloud corresponding to a nonzero latent heat $\DeltaQ =(250\, \textrm{MeV})^4$ at the matching point of the two monotropes, $\mu=\mu_c$. The black curves correspond to a number of individual representative EoSs.} 
\label{fig8}
\end{figure}

As stiffer EoSs produce heavier stars, the maximal speed of sound predicted by a given EoS is naturally correlated with the corresponding maximal mass. We find that in order to be able to fulfill the $2M_{\odot}$ constraint, this maximal value has to satisfy $c_s^2 > 0.55$. We suspect, however, that this constraint may be overly restrictive, as our interpolating polytropes typically predict a discontinuous $c_s^2$ that is peaked around $\mu=\mu_c$.

\subsection{Robustness of the results}
In deriving the results presented above we made two in principle \textit{ad hoc} assumptions, whose effect on the obtained EoS and $M-R$ relations we now proceed to study. First, one can naturally extend the number of interpolating monotropes to three (or more), allowing for further freedom in the behavior of the EoS between the low and high density regimes. Second, the assumption of a smooth matching of the two monotropes can (and should) be relaxed by allowing for a jump in the number density at the matching point, corresponding to a first-order phase transition between the nuclear and quark matter phases.

Starting with the number of monotropes, we generated a large set of tritropic EoSs with randomly chosen matching points and polytropic indices, which fulfill the smoothness and subluminality constraints. Fig.~\ref{fig7} depicts the effect of this extra freedom on the EoS: we observe that, as expected, the addition of a third segment somewhat increases the region of allowed EoSs, but the effect is small in comparison with the other uncertainties of the calculation. From here, we conclude that using bitropic EoSs should suffice for our purposes.

\begin{figure}
\includegraphics[width=0.45\textwidth]{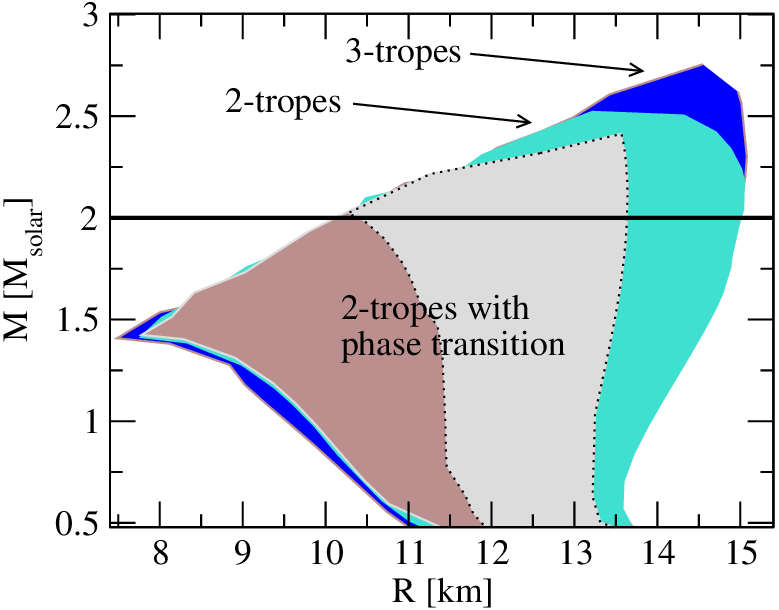}
\caption{The $M$-$R$ clouds corresponding to tritropic and first-order phase transition EoSs, shown together with our original result from Fig.~\ref{fig3}.} 
\label{fig9}
\end{figure}

To investigate the effect of a first-order phase transition at $\mu_c$, we next relaxed the smoothness condition in our matching of the two monotropes. Keeping the pressure continuous but allowing for a `latent heat' $\DeltaQ \equiv \mu_c\Delta n$ of the order of the QCD scale, we first fixed $\DeltaQ = (250\,\textrm{MeV})^4$ and proceeded to find solutions for $\gamma_1$ and $\gamma_2$ that would lead to a consistent EoS. This led to the rather restricted values $\gamma_1 \in [2.23,4.03]$ and $\gamma_2 \in [1,1.5]$, for which the transition point was always found to lie within the interval $\mu_c \in [1.4,2.1]\,$GeV. The corresponding region of allowed EoSs, depicted in Fig.~\ref{fig8}, was found to be somewhat smaller than in the case with smooth matching. From this (as well as similar calculations performed for $\DeltaQ=(175\,\rm{MeV})^4$ and $\DeltaQ=(225\,\rm{MeV})^4$), we conclude that the assumption of smooth matching made in the previous section was in fact justified when searching for the least 
restrictive bounds for the EoS.

For EoSs displaying a phase transition, one can also estimate the amount of quark matter in the cores of the stars. This is seen from Fig.~\ref{fig6}, which shows the relation between the maximal chemical potential reached at the center of a maximally massive star
$\mu_{\rm center}$ and the critical (matching) chemical potential $\mu_c$. We see that all EoSs that fulfill the mass constraint lie above the $\mu_{\rm center}>\mu_c$ line, and are therefore able to support stars with quark matter cores. However, the stronger the transition is, the smaller the window for quark matter: for $\Delta Q =(250 \rm{GeV})^4$, there is practically no quark matter left in the cores of the stars.

In Fig.~\ref{fig9}, we finally show the effect of the third monotrope and a nonzero latent heat on the obtained $M-R$ clouds. In particular, we see from here that allowing for a tritropic interpolation does not have a large impact on the $M-R$ plot: the most important change is simply the shift of the maximal mass star to $\{M_{\rm max},R\} = \{2.75M_{\odot},14.6{\rm km}\}$. A more complete analysis of the case of a first-order phase transition has been recently performed by \cite{Alford:2013aca}. In this reference, the authors in particular consider all possible branching cases, including twin star 
configurations, which we have completely omitted in our work.

\begin{figure}
\includegraphics[width=0.45\textwidth]{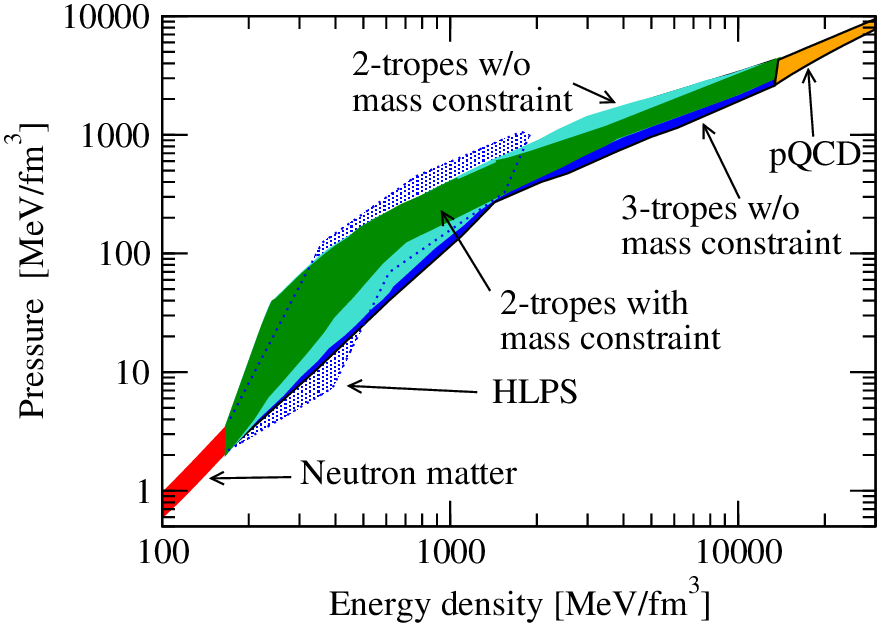}
\caption{A comparison of our EoSs with those of \cite{Hebeler:2013nza}, labeled HLPS in the figure. As is clear from the sizes of the green and light blue regions, corresponding respectively to our bitropic EoSs and the HLPS results (with the two solar mass constraint implemented in both), the high-density constraint significantly shrinks the allowed range of EoSs.}
\label{fig10}
\end{figure}


\section{Conclusions and summary}

In the paper at hand, we have constructed a novel scheme for determining the EoS of compact star matter that involves an interpolation between the regimes of low-energy chiral effective theory and high-density perturbative QCD. These two limiting results are truly robust within their ranges of applicability, as they represent controlled calculations in the fundamental theory of the strong interactions. Our work on the other hand constitutes the first ever attempt to take constraints from both sides on equal footing when determining the EoS between these limits. We have demonstrated that this leads to important new constraints on the properties of compact star matter on a wide density range, and thus even for stars containing only hadronic matter.

The strictness of the constraints placed on the stellar EoS by its high-density limit can be understood through the tension between the softness of the perturbative EoS and the stiffness required by the confirmed existence of a two solar mass compact star. For the two interpolating monotropes we employ in our calculation, this translates into a significant difference between the respective polytropic indices: While the first one needs to be rather stiff, with $\gamma_1>2.86$, the latter must be considerably softer, $1<\gamma_2<1.5$. Although the polytropes themselves of course do not carry information about the underlying microphysics, such a strong shift in the polytropic index might be interpreted as a sign of the effective degrees of freedom of the system changing from hadronic to denconfined ones. 

The effect of the high density constraint is perhaps best illustrated in Fig.~\ref{fig10}, which displays our EoS band in the form of energy density vs.~pressure, plotted together with the previous prediction of \cite{Hebeler:2013nza}, dubbed HLPS. The latter work applied the same low-density EoS we did and took into account the two solar mass constraint, but did not require the result to approach the pQCD EoS at large densities. As expected, the main difference between the two results is seen in the HLPS cloud containing somewhat softer EoSs at low density and stiffer ones at high density.

The rather narrow EoS band that results from our interpolation naturally corresponds to a well defined region in the mass-radius diagram of compact stars. For a $1.4M_\odot$ neutron star, the radii we obtain range between 11 and 14.5$\,$km, while the radius of a $2M_\odot$ pulsar lies within $R\approx 10-15\,$km. Interestingly, we do not find configurations with masses above $2.75M_\odot$ (for bitropic interpolation the maximal mass is $2.5M_\odot$). This conclusion is in contrast with what has been found before without the high-density constraint; see e.g.~\cite{Hebeler:2013nza}, where stars with masses up to $3M_\odot$ were discovered.

For the convenience of the reader, we finally provide three representative EoSs in a tabulated form at the end of this paper. These EoSs are all subluminal, able to sustain a two solar mass
star, and maximally different from each other. Of them, EoS I gives the minimal radius, EoS II the maximal mass and EoS III the maximal radius for our compact stars.

In conclusion, we find it remarkable, how the properties of quark matter at asymptotically high densities can be seen to have such a strong impact on the structure of compact stars at much lower energies. As we have highlighted in Fig.~\ref{fig1}, this fact appears to make it possible to largely bridge the gap between the respective EoSs of low-density nuclear matter and high-density (perturbative) quark matter.


\section*{Acknowledgments} 
The authors would like to thank Kai Hebeler for useful discussions. AK and ESF in addition acknowledge the Helsinki Institute of Physics, AK and AV HIC4FAIR, and ESF the Theory Division of CERN for hospitality and financial support. The work of ESF was supported by the Helmholtz International Center for FAIR within the framework of the LOEWE program (Landesoffensive zur Entwicklung Wissenschaftlich-\"Okonomischer Exzellenz) launched by the State of Hesse, and that of AV by the Academy of Finland, grant \# 266185.

\clearpage

\begin{table}
$$\begin{tabular}{c|ccccc|cc}
\#	&Nucl. & $\mu_c$&$X$	&$\gamma_1$	&$\gamma_2$&	$M_{\rm max}$	& $\mu_{\rm max}$\\
\hline
I	&s	&1.65	&1.2	&3.192	&1.024	&	2.03		&1.78\\
II	&h	&1.35	&1.2	&4.021	&1.195	&	2.44  	&1.54\\
III	&h	&1.125	&1.9	&7.368	&1.415	&	2.24   	&1.36\\
\hline
4	&h	&1.125	&4.0	&4.585	&1.483	&	1.95	        &1.36 \\
5	&h	&1.35	&1.4	&3.440	&1.258	&	2.20		&1.54\\
6	&h	&1.35	&4.0	&2.698	&1.407	&	1.75		&1.53\\
7	&h	&1.65	&1.2	&2.865	&1.051	&	2.01		&1.77\\
8	&h	&1.65	&2.0	&2.494	&1.206	&	1.67		&1.74\\
9	&h	&1.65	&4.0	&2.370	&1.280	&	1.54		&1.72\\
10	&h	&1.95	&2.0	&2.335	&1.102	&	1.50		&1.76\\
11	&h	&1.95	&4.0	&2.253	&1.109	&	1.41		&1.71\\
12	&s	&1.125	&4.0	&5.322	&1.474	&	1.97		&1.36\\
13	&s	&1.125	&2.1	&7.439	&1.422	&	2.19		&1.36\\
14	&s	&1.35	&1.3	&4.136	&1.215	&	2.31		&1.56\\
15	&s	&1.35	&1.6	&3.606	&1.278	&	2.08		&1.54\\
16	&s	&1.35	&4.0	&3.043	&1.382	&	1.76		&1.54\\
17	&s	&1.65	&2.0	&2.771	&1.167	&	1.69		&1.77\\
18	&s	&1.65	&4.0	&2.630	&1.234	&	1.56		&1.76\\
19	&s	&1.95	&2.8	&2.517	&1.001	&	1.45		&1.87\\
20	&s	&1.95	&4.0	&2.481	&1.028	&	1.41		&1.86
\end{tabular}
$$
\caption{Parameter values for a representative set of EoSs resulting from bitropic interpolation. The letters ${\rm s}$ (soft) and ${\rm h}$ (hard) in the column ``Nucl.'' refer to the use of Eqs.~(\ref{Nucl:s}) and (\ref{Nucl:h}) at $n=n^{\rm begin}$, respectively. The chemical potentials are given in ${\rm GeV}$ and the maximal masses in solar masses.}
\label{table1} 
\end{table}

\begin{table}
$$
\begin{array}{c||cccc|cc}
n/n_0	&	P &	E	&	\mu_B &c_s^2&	R &	M/M_{\odot}\\
\hline
\hline
 1.1 & 2.163 & 167.8 & 0.9657 & 0.041 & 22.2 & 0.144 \\
 1.3 & 3.687 & 198.8 & 0.9736 & 0.058 & 18.1 & 0.168 \\
 1.5 & 5.822 & 230.1 & 0.9831 & 0.079 & 14.3 & 0.231 \\
 1.7 & 8.681 & 261.8 & 0.9943 & 0.10 & 12.6 & 0.322 \\
 1.9 & 12.38 & 293.8 & 1.007 & 0.13 & 12.0 & 0.415 \\
 2.1 & 17.04 & 326.2 & 1.022 & 0.16 & 11.6 & 0.526 \\
 2.3 & 22.79 & 359.2 & 1.038 & 0.19 & 11.5 & 0.652 \\
 2.5 & 29.74 & 392.7 & 1.056 & 0.22 & 11.4 & 0.768 \\
 2.7 & 38.02 & 426.8 & 1.076 & 0.26 & 11.4 & 0.875 \\
 2.9 & 47.76 & 461.6 & 1.098 & 0.30 & 11.4 & 0.989 \\
 3.1 & 59.09 & 497.1 & 1.121 & 0.34 & 11.3 & 1.11 \\
 3.3 & 72.14 & 533.3 & 1.147 & 0.38 & 11.3 & 1.22 \\
 3.5 & 87.05 & 570.5 & 1.174 & 0.42 & 11.3 & 1.32 \\
 3.7 & 103.9 & 608.5 & 1.203 & 0.47 & 11.3 & 1.42 \\
 3.9 & 123.0 & 647.5 & 1.235 & 0.51 & 11.2 & 1.50 \\
 4.1 & 144.3 & 687.6 & 1.268 & 0.55 & 11.2 & 1.58 \\
 4.3 & 167.9 & 728.7 & 1.303 & 0.60 & 11.1 & 1.66 \\
 4.5 & 194.2 & 771.0 & 1.341 & 0.64 & 11.0 & 1.72 \\
 4.7 & 223.1 & 814.5 & 1.380 & 0.69 & 10.9 & 1.78 \\
 4.9 & 254.8 & 859.3 & 1.421 & 0.73 & 10.9 & 1.83 \\
 5.1 & 289.5 & 905.5 & 1.464 & 0.77 & 10.8 & 1.88 \\
 5.3 & 327.4 & 953.1 & 1.510 & 0.82 & 10.7 & 1.92 \\
 5.5 & 368.5 & 1002. & 1.558 & 0.86 & 10.6 & 1.95 \\
 5.7 & 413.0 & 1053. & 1.607 & 0.90 & 10.5 & 1.98 \\
 \hline
 5.9 & 455.3 & 1105. & 1.653 & 0.30 & 10.5 & 2.01 \\
 6.1 & 471.1 & 1158. & 1.669 & 0.30 & 10.4 & 2.01 \\
 6.3 & 486.9 & 1212. & 1.685 & 0.29 & 10.4 & 2.02 \\
 6.5 & 502.8 & 1266. & 1.701 & 0.29 & 10.4 & 2.02 \\
 6.7 & 518.6 & 1321. & 1.716 & 0.29 & 10.4 & 2.03 \\
 6.9 & 534.5 & 1376. & 1.730 & 0.29 & 10.3 & 2.03 \\
 7.1 & 550.3 & 1431. & 1.744 & 0.28 & 10.3 & 2.03 \\
 7.3 & 566.2 & 1487. & 1.758 & 0.28 & 10.3 & 2.03 \\
 7.5 & 582.1 & 1544. & 1.772 & 0.28 & 10.3 & 2.03
\end{array}
$$
\caption{The representative equation of state I. The baryon number density $n$ is given in units of the saturation density $n_0=0.16\,\rm{fm}^3$, while the pressure $P$ and the energy density $E$ are given in $\rm{ MeV/fm}^3$. $R$ (in ${\rm km}$) and $M$
(in solar masses) stand for the radius and mass of a star with central density $n$, while the solid horizontal line indicates the transition between the two monotropes.}
\label{table2}
\end{table}

\begin{table}
$$
\begin{array}{c||cccc|cc}
n/n_0	&	P &	E	&	\mu	&c_s^2&	R &	M/M_\odot\\
\hline
\hline
 1.1 & 3.542 & 168.5 & 0.9775 & 0.083 & 15.7 & 0.259 \\
 1.3 & 6.934 & 200.0 & 0.9951 & 0.13 & 13.9 & 0.369 \\
 1.5 & 12.33 & 232.3 & 1.019 & 0.20 & 13.2 & 0.590 \\
 1.7 & 20.39 & 265.3 & 1.051 & 0.29 & 13.1 & 0.880 \\
 1.9 & 31.89 & 299.6 & 1.090 & 0.39 & 13.3 & 1.18 \\
 2.1 & 47.70 & 335.2 & 1.140 & 0.50 & 13.5 & 1.50 \\
 2.3 & 68.76 & 372.6 & 1.199 & 0.63 & 13.6 & 1.79 \\
 2.5 & 96.16 & 412.1 & 1.271 & 0.76 & 13.6 & 2.06 \\
 \hline
 2.7 & 129.7 & 454.1 & 1.351 & 0.27 & 13.6 & 2.28 \\
 2.9 & 141.2 & 497.7 & 1.377 & 0.26 & 13.6 & 2.33 \\
 3.1 & 153.0 & 542.2 & 1.402 & 0.26 & 13.6 & 2.37 \\
 3.3 & 164.8 & 587.4 & 1.425 & 0.26 & 13.5 & 2.40 \\
 3.5 & 176.8 & 633.4 & 1.447 & 0.26 & 13.5 & 2.42 \\
 3.7 & 189.0 & 680.0 & 1.468 & 0.26 & 13.4 & 2.43 \\
 3.9 & 201.2 & 727.3 & 1.488 & 0.26 & 13.4 & 2.44 \\
 4.1 & 213.6 & 775.2 & 1.507 & 0.26 & 13.3 & 2.45 \\
 4.3 & 226.1 & 823.8 & 1.526 & 0.26 & 13.3 & 2.45 \\
 4.5 & 238.7 & 872.9 & 1.544 & 0.26 & 13.2 & 2.45

\end{array}
$$
\caption{The representative equation of state II, with conventions as explained in Table \ref{table2}.}
\end{table}

\begin{table}
$$
\begin{array}{c||cccc|cc}
n/n_0	&	P &	E	&	\mu	&c_s^2&	R &	M\\
\hline
\hline
  1.1 & 3.542 & 168.5 & 0.9775 & 0.15 & 15.7 & 0.259 \\
 1.3 & 12.13 & 200.4 & 1.022 & 0.42 & 13.4 & 0.670 \\
 1.5 & 34.81 & 234.5 & 1.122 & 0.95 & 14.7 & 1.66 \\
 \hline
 1.7 & 42.24 & 270.9 & 1.151 & 0.19 & 14.9 & 1.85 \\
 1.9 & 49.44 & 308.1 & 1.176 & 0.20 & 15.0 & 1.99 \\
 2.1 & 56.96 & 346.2 & 1.200 & 0.20 & 15.0 & 2.08 \\
 2.3 & 64.79 & 384.9 & 1.222 & 0.20 & 15.0 & 2.13 \\
 2.5 & 72.90 & 424.4 & 1.243 & 0.21 & 14.9 & 2.17 \\
 2.7 & 81.29 & 464.5 & 1.263 & 0.21 & 14.9 & 2.20 \\
 2.9 & 89.94 & 505.2 & 1.283 & 0.21 & 14.8 & 2.22 \\
 3.1 & 98.84 & 546.5 & 1.301 & 0.22 & 14.7 & 2.23 \\
 3.3 & 108.0 & 588.5 & 1.319 & 0.22 & 14.6 & 2.24 \\
 3.5 & 117.4 & 631.0 & 1.336 & 0.22 & 14.5 & 2.24 \\
 3.7 & 127.0 & 674.0 & 1.353 & 0.22 & 14.4 & 2.24 \\
 3.9 & 136.8 & 717.5 & 1.369 & 0.23 & 14.3 & 2.24 
\end{array}
$$
\caption{The representative equation of state III, with conventions as explained in Table \ref{table2}.}
\label{table4}
\end{table}


\end{document}